\documentclass[12pt]{iopart}
\usepackage[section]{placeins}
\expandafter\let\csname equation*\endcsname\relax
\expandafter\let\csname endequation*\endcsname\relax
\usepackage{amsmath,amsfonts,amssymb}
\usepackage{pb-diagram}
\usepackage[english]{babel}
\usepackage{tikz}

\let\p\partial

\let\ds\displaystyle

\begin{document}
\title{Quad-equations and auto-B\"acklund transformations of NLS-type systems}

\author{D.K. Demskoi}
\address{
School of Computing and Mathematics,\\ Charles Sturt University, NSW 2678, Australia 
}

\begin{abstract}
Treating an integrable quad-equation along with its 
two generalised symmetries as a compatible system allows one to construct
an auto-B\"acklund transformation for solutions of the related 
NLS-type system. A fixed periodic reduction of the quad-equation
yields a quasi-periodic reduction of its generalised symmetries 
that turn them into differential constraints compatible with the NLS-type system.
\end{abstract}
 
\section{Introduction}
Integrable differential-difference equations with one continuous and one discrete variable,
subsequently referred to as {\it chains}, are known to be closely connected with (systems of) integrable 
partial differential equations (PDEs). In particular, many integrable chains can be interpreted as 
B\"acklund transformations of some PDEs \cite{levi}. The integrability of a chain assumes 
the existence of a formal recursion operator and infinitely many commuting flows. This property
has been used to classify both integrable chains and PDEs \cite{AdlShYa}. A pair of commuting flows
from the same hierarchy is called {\it compatible}.
Shabat and Yamilov demonstrated that a pair of compatible chains with some restrictions on their form 
always yields a system of PDEs through the construction often referred to as elimination of shifts 
\cite{ShY}. A by-product of this construction is an invertible auto-transformation of the resulting system of PDEs.

As far as construction of exact solutions is concerned, a more
important class of transformations is 
non-invertible auto-transformations
containing an arbitrary parameter (auto-B\"acklund transformations). 
A direct calculation of such
transformations is a tedious task. The knowledge of other 
structures associated with integrability, e.g. a Lax pair or Painlev\'e structure, may significantly 
simplify the calculation of such transformations \cite{weiss}.
In this paper we show how an auto-B\"acklund transformation
can be constructed when a system of PDEs is obtainable 
through the elimination of shifts from a compatible system of two integrable chains.
The necessary ingredient in this construction is that
the chains should represent the generalised symmetries 
of an integrable quad-equation.

To illustrate the idea we consider the integrable chain 
\begin{equation}
\ds\p_x u_{k,l}= \frac{1}{u_{k+1,l}-u_{k-1,l}},
\label{S_1}
\end{equation} 
where $u_{k,l}=u(k,l;x,y)$ is a function that simultaneously depends on discrete 
and continuous variables: $(k,l)\in \mathbb{Z}^2,\, (t,x) \in \mathbb{C}^2$.
Throughout the article the subscripts $k$ and $l$ indicate dependence on
discrete variables, while the subscripts $t$ and $x$ indicate partial derivatives.
{Equation (\ref{S_1}) is related to the famous Volterra equation 
$$
\ds\p_x w_{k,l}= w_{k,l}(w_{k+1,l}-w_{k-1,l})
$$
via the substitution \cite{yamcherd}
$$
w_{k,l}=-\frac{1}{(u_{k+1,l}-u_{k-1,l})(u_{k+2,l}-u_{k,l})}.
$$
The complete classification of the Volterra-type equations can be found in \cite{Yam}.
}

{
The simplest commuting flow, i.e. an equation $\p_t u_{k,l}=G$ of the lowest order  
that satisfies $\p_t\p_x u_{k,l}=\p_x\p_t u_{k,l}$, of (\ref{S_1}) 
is given by
\begin{equation}
\p_t u_{k,l}=\frac{u_{k+2,l}-u_{k-2,l}}{( u_{k+1,l}-u_{k-1,l})^{2}(u_{k+2,l}-u_{k,l} )( u_{k,l}-u_{k-2,l})}. 
\label{S_2}
\end{equation} 
It can be computed by using the standard tools, such as master symmetry \cite{yamcherd}
or recursion operator \cite{mikhxen}. On the other hand the whole hierarchy of (\ref{S_1})
can be represented by a single formula (see formula (9) of \cite{svinin}).}
Note that neither of chains (\ref{S_1}) or (\ref{S_2}) depends on shifts with respect to variable $l$.
Nevertheless it is indicated here in order to make possible a connection with 
a quad-equation (see below).

{In order to obtain a system of PDEs satisfied by $u_{k,l}$ and $u_{k+1,l}$, we use (\ref{S_1}) 
and its shifted versions to express variables $u_{k-2,l},\, u_{k-1,l}$ and $u_{k+2,l}$:
\begin{equation}
u_{k-2,l} = u_{k,l } -\frac{1}{\p_x u_{k-1,l }}, \ \	u_{k-1,l} = u_{k+1,l } -\frac{1}{\p_x u_{k,l }}, \ \  u_{k+2,l } = u_{k,l}+\frac{1}{\p_x u_{k+1,l }}.
	\label{shfts}
\end{equation}}
The substitution of (\ref{shfts}) into (\ref{S_2}) yields the derivative NLS system \cite{dnls} in the potential form:
\begin{equation}
\begin{array}{l}
u_t=u_{xx}+2 u_x^2 v_x, \\[1mm]
v_t=-v_{xx}+2 v_x^2 u_x,
\end{array}
\label{dNLS}
\end{equation}
where
$u_{k,l}=u,\,  u_{k+1,l}=v.$
{
The shifts along chain (\ref{S_1})
$$(u_{k,l},u_{k+1,l})\to (u_{k+1,l},u_{k+2,l}) ,\ \ (u_{k-1,l},u_{k,l})\to (u_{k,l},u_{k+1,l}), $$
can now be interpreted as the auto-transformation of (\ref{dNLS})
\begin{equation}
	\left(\begin{array}{c}u \\ v \end{array}\right)\to
	\left(\begin{array}{c}v \\ u+1/v_x \end{array}\right)
	\label{upshift}
\end{equation}
and its inverse
\begin{equation}
	\left(\begin{array}{c}u \\ v \end{array}\right)\to
	\left(\begin{array}{c}v-1/u_x \\ u \end{array}\right)  
	\label{downshift}
\end{equation}
correspondingly.
}

It is known that integrable quad-equations
possess hierarchies of generalised symmetries (see e.g. \cite{hydon, leviyamsym}). For instance, the 
hierarchy of equations (\ref{S_1}) and (\ref{S_2}) is related
to the quad-equation \begin{equation}
	(u_{k,l}-u_{k+1,l+1})(u_{k+1,l}-u_{k,l+1})-\lambda+\mu=0,
	\label{h1}
\end{equation}
where $\lambda,\mu=\mbox{const}$.
This equation is often referred to as $H_1$ due to the labeling it received in the 
classification \cite{ABS} of equations consistent around the cube.
{The $H_1$ equation is also well known in the context of the potential 
KdV equation where
it serves as a superposition formula for solutions related by the auto-B\"acklund transformation \cite{wahlq}. 
Moreover, equation (\ref{h1}) reduces to pKdV in the continuum limit \cite{qncl,qncl2}.}
This example therefore highlights the link between the classes 
of NLS and KdV-type equations. 

In what follows we are concerned with implications of the mentioned connection 
between integrable chains, NLS-type systems and quad-equations. 
We show that it automatically yields an auto-B\"acklund
transformation for the related NLS-system. {A formula
of superposition can then be derived from the assumption of commutativity
of the auto-B\"acklund transformations.} 
In general the compatibility of a PDE and a superposition formula needs to 
be verified separately, and is not always guaranteed.
One of the corollaries of the presented construction is that
a traveling wave reduction of an integrable quad-equation generates 
the quasi-periodic closure of the related chains, which turn them into 
differential constraints compatible with the NLS-type system.
\section{Auto-B\"acklund transformations of NLS-type systems}
The statement that (\ref{S_1}) and (\ref{S_2}) are generalised symmetries of (\ref{h1}) 
implies that the relations
\begin{equation}
	\p_t F=0, \ \ \p_x F=0,
	\label{ident1}
\end{equation}
where $F$ is the left hand side of (\ref{h1}), are identically satisfied on solutions 
of the system consisting of (\ref{S_1}), (\ref{S_2}) and (\ref{h1}). In other words, (\ref{ident1}) 
become identities
when partial derivatives are eliminated by using (\ref{S_1}) and (\ref{S_2}), 
and mixed shifts by using (\ref{h1}).

Note that due to the symmetry $(k,l)\to (l,k)$, equation 
(\ref{h1}) possesses the generalised symmetries of the form (\ref{S_1}) and (\ref{S_2}), where $k$ and 
$l$ are interchanged. However, the corresponding system of PDEs will still be 
the same (potential dNLS). The construction being considered here can be applied to 
non-symmetrical quad-equations to show that one quad-equation can generate auto-B\"acklund
transformations for two different NLS-type systems. However, for the sake of simplicity we will consider
only the example of the $H_1$ equation.

Since equations (\ref{S_1}) and (\ref{S_2}) do not involve shifts with respect to the variable 
$l$, the quantities 
\begin{equation*}
	p=u_{k,l+1},\ \  q=u_{k+1,l+1}  
	\label{pq}
\end{equation*}
must satisfy a system of form (\ref{dNLS}) with $(u,v)$ being replaced by $(p,q)$:
\begin{equation}
\begin{array}{l}
p_t=p_{xx}+2 p_x^2 q_x, \\[1mm]
q_t=-q_{xx}+2 q_x^2 p_x.
\label{dnlsp}
\end{array}
\end{equation}
This observation implies that quad-equation (\ref{h1}) when re-written as 
\begin{equation}
(u-q)(v-p)=\kappa, 
\label{backl1}
\end{equation}
where $\kappa=\lambda-\mu$,
is a part of a certain auto-transformation for the potential dNLS system. 
Importantly, the constant $\kappa$ is not present
in (\ref{dnlsp}); hence it can play the role of the B\"acklund parameter.
Another part of the auto-transformation can be found the following way.

Consider the up- and down-shifted versions of (\ref{h1}):
\begin{equation}
(u_{k+1,l}-u_{k+2,l+1})(u_{k+2,l}-u_{k+1,l+1})=\kappa,
\label{upshifted}
\end{equation}
\begin{equation}
(u_{k-1,l}-u_{k,l+1})(u_{k,l}-u_{k-1,l+1})=\kappa.
\label{downshifted} 
\end{equation} 
It follows from (\ref{S_1}) that
$$
\ds u_{k+2,l}=u+\frac{1}{v_x}, \ \ u_{k+2,l+1}=p+\frac{1}{q_x},\ \ 
\ds u_{k-1,l}=v-\frac{1}{u_x}, \ \ u_{k-1,l+1}=q-\frac{1}{p_x}.
$$
Substituting these expressions into (\ref{upshifted}) and (\ref{downshifted}) we obtain
the additional relations
\begin{equation}
\left(v-p-\tfrac{1}{q_x}\right)\left(u-q+\tfrac{1}{v_x}\right)=\kappa, 
\label{dubackl}
\end{equation}
\begin{equation}
\left(v-p-\tfrac{1}{u_x}\right)\left(u-q+\tfrac{1}{p_x}\right)=\kappa. 
\label{ddbackl}
\end{equation}
{One can verify 
that the combination of (\ref{backl1}) and (\ref{dubackl}) implies formula (\ref{ddbackl}).
Therefore any combination of two relations from the list of (\ref{backl1}), 
(\ref{dubackl}) and (\ref{ddbackl}) constitutes an auto-B\"acklund transformation for (\ref{dNLS}).}
The analogous transformations for the dNLS system were previously constructed in \cite{kundu,steudel}
by using different approaches.
\subsection*{Superposition formula and construction of solutions}
Now we turn to constructing a superposition formula based on the 
auto-B\"acklund transformation found previously, i.e. the combination of relations 
(\ref{backl1}) and (\ref{dubackl}). To this end we look at implications of commutativity of a 
few transformations (\ref{backl1}) which can be schematically represented by 
the Bianchi diagram:
\begin{equation*}
\begin{diagram}
\node{(m,n)}\arrow{e,t}{\kappa}\node{(r,s)} \\
\node{(u,v)}\arrow{e,b}{\kappa}\arrow{n,l}{\nu}\node{(p,q)}\arrow{n,r}{\nu}
\end{diagram}
\label{diag}
\end{equation*}
The relation (\ref{dubackl}) is used to obtain the new solution from a seed solution. 
The diagram yields the following relations
$$
\begin{array}{l}
(u-q)(v-p)=\kappa, \\
(u-n)(v-m)=\nu, 
\end{array} \ \ 
\begin{array}{l}
(p-s)(q-r)=\nu, \\
(m-s)(n-r)=\kappa
\end{array}
$$
which in turn give rise to the possible expressions for $r$ and $s$: 
{
\begin{equation}
\ds r=u+\frac{\kappa-\nu}{p-m}, \ \ 
\ds s= v+\frac{\kappa-\nu}{q-n}
\label{sf1}
\end{equation}}
and
\begin{equation}
	r = n+q-u, \ \ s = v-\frac{\nu}{u-n}-\frac{\kappa}{u-q}.
	\label{sf2}
\end{equation}
One can check that the second relation is not compatible with the dNLS system, whereas the first one is!
The compatibility is verified by differentiating (\ref{sf1}) (or  (\ref{sf2})) with respect to the time variable and then
making use of the potential dNLS system itself, and also of (\ref{backl1}), (\ref{dubackl}) and (\ref{sf1}) (or (\ref{sf2})).
{Obviously (\ref{sf1}) is nothing but the two copies of the standard potential KdV superposition 
formula relating the corresponding components in the Bianchi diagram. Note that (\ref{sf1}) is not the only possible
form of the superposition formula since $m$ and $p$ could be eliminated from the formula.
}

By iterating formula (\ref{sf1}) we obtain rational expressions in terms of a seed solution and the solution obtained
through the dNLS system (\ref{dnlsp}), (\ref{backl1}) and (\ref{dubackl}). 

{{\bf Example.} If we start with the exponential solution
$$
u=\exp(x-t),\ \ v=\exp(-x+t),
$$
then it follows that $q$ satisfies the system
\begin{equation}
	q_x = \frac{q(1-q v)}{\kappa},\ \ q_t = \frac{(1+\kappa) q^2 v-q}{\kappa^2}
	\label{toint}
\end{equation}
while $p$ is given explicitly by
$$
p = v+\frac{\kappa}{q}-\frac{1}{q_x}.
$$
Integrating equations (\ref{toint}), we obtain
$$
q=\frac{1-\kappa}{v+c\exp(-\tfrac{x}{\kappa}+\tfrac{t}{\kappa^2})},
$$
where $c$ is the constant of integration.}

{A more intricate solution is then obtained through superposition formula (\ref{sf1}). Note that expressions for
$m$ and $n$ coincide with $p$ and $q$ correspondingly, where the parameter $\kappa$ is replaced by $\nu$.
A common feature of the solutions obtained from the exponential seed solution is that the individual
components grow/decay exponentially while their product
has the shape of a multi-soliton solution.
Such solutions are called {\it dissipatons} \cite{pash}. 
For instance, for the values of parameters $\kappa=2,\, \nu=1/2,\, c=1$ the plot for the product of $r$ and  $s$ 
is} \newpage

\begin{figure*}[h]  
 \begin{center}  
 \includegraphics[angle=0,scale=0.12]{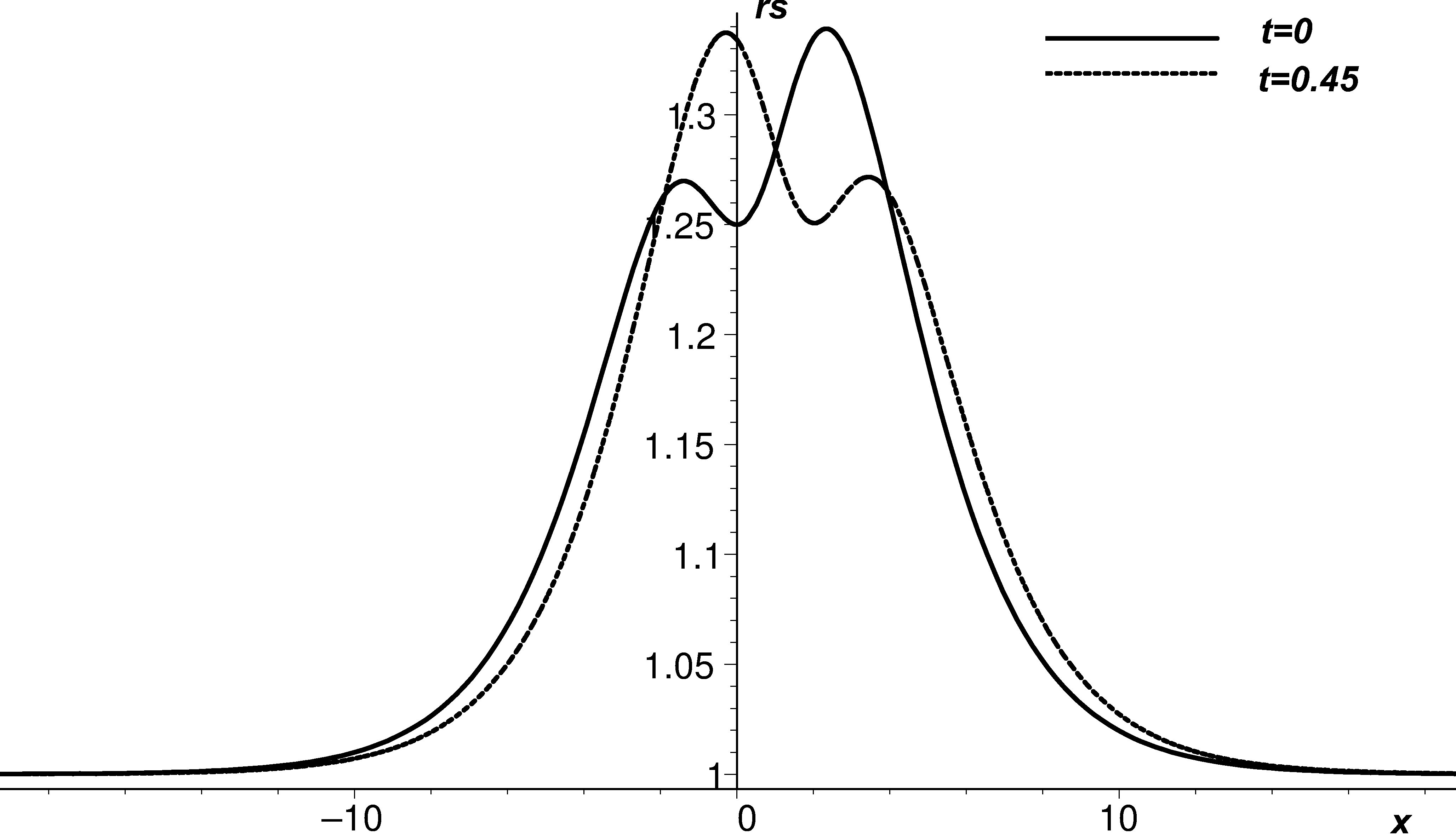}  
 \end{center}  
 \end{figure*}

{\bf Remark. }%
The fact that equation (\ref{h1}) serves two different hierarchies suggests the presence of a common
member in the KdV and potential dNLS hierarchies.
Indeed, the hierarchy of chains (\ref{S_1}) and (\ref{S_2})
also contains the ``negative'' flow
\begin{equation}
	\p_z u_{k,l} = -\p_z u_{k+1,l}+(u_{k,l}-u_{k+1,l})^2+\lambda.
	\label{sm1} 
\end{equation}
By differentiating (\ref{S_1}) and (\ref{sm1}) with respect to $z$  and $x$ correspondingly
and then eliminating shifts from the obtained expressions, we get the hyperbolic system
\begin{equation}
	\begin{array}{l}
	u_{xz} = 2(u-v)u_x+1, \\[1mm]
	v_{xz} = -2(u-v)v_x-1.
	\end{array}
	\label{hyp}
\end{equation}
It is not difficult to verify that (\ref{hyp}) commutes with the potential dNLS system.
On the other hand, the compatibility of chains (\ref{S_1}) and (\ref{sm1})
can be written as one scalar equation \cite{AdlSh}
\begin{equation}
	u_{xzz}= \frac 12\frac{u_{xz}^2-1}{u_x} +2 u_x (2u_z -\lambda),
	\label{kh}
\end{equation}
which commutes with the potential KdV equation 
$$
u_t=u_{zzz}-6 u_z^2.
$$
\subsection*{Reductions}
Here we discuss the connections of periodic reductions of quad-equations and
quasi-periodic closures of the integrable chains. In fact we could have come 
to the same construction of auto-B\"acklund transformations by considering the 
reductions $u_{k,l}\to u_{\alpha k+\beta l}$, {where $\alpha$ and $\beta$ are some integers}, 
which induce the periodicity constraint 
$u_{k,l}=u_{k-\beta,l+\alpha}$. The simplest reduction of this type is when $\alpha=1$. 
This reduction, being applied to equation (\ref{h1}), brings it to the form 
\begin{equation}
(u_{k}-u_{k+\beta+1})(u_{k+1}-u_{k+\beta})=\kappa.
\label{h1r}
\end{equation}
It is important that chains (\ref{S_1}) and (\ref{S_2}) survive this reduction 
for an arbitrary $\beta$ and become the symmetries of (\ref{h1r}) upon the substitution
$u_{k+i,l}\to u_{k+i}$.
Moreover, the same procedure of elimination of shifts yields the potential
dNLS system with unknowns $u_k=u$ and $u_{k+1}=v$.

Since $\beta$ is arbitrary, the quantities 
$$u_{k+\beta}=p,\ \ u_{k+\beta+1}=q$$ 
should be treated as algebraically independent from $u_{k}$  and $u_{k+1}$. 
Thus equation (\ref{h1r}) yields the auto-transformation
$$
(u-q)(v-p)=\kappa
$$
of (\ref{dNLS}) into itself. Relations (\ref{dubackl}) and (\ref{ddbackl}) can be derived 
in exactly the same way as before.

In the case when $\beta$ is fixed, the quantities $u_{k+\beta}$ and $u_{k+\beta+1}$
can no longer be treated as independent because we can express them in terms 
of $u_k$ and $u_{k+1}$ by using the reduction of (\ref{S_1}). As a result 
we obtain a differential 
constraint in the form of a dynamical system compatible with 
the potential dNLS equation. On the other hand, the periodicity constraint
transforms the quad-equation into an ordinary difference equation which can
be interpreted as a mapping acting in a finite-dimensional space. By construction 
this mapping will preserve the differential constraint.

{\bf Example.} Consider the case $\alpha=1,\, \beta=2$. Equation (\ref{h1}) turns into the 
ordinary difference equation
\begin{equation}
(u_{k}-u_{k+3})(u_{k+1}-u_{k+2})=\kappa,
\label{h1r3}
\end{equation}
while chain (\ref{S_1}) becomes
\begin{equation}
\ds\p_x u_{k}= \frac{1}{u_{k+1}-u_{k-1}}.
\label{s1r}
\end{equation}
Writing (\ref{s1r}) for $k=0\dots 2$ and eliminating $u_{-1}$ and  $u_3$ using (\ref{h1r3}), we obtain
the system
\begin{equation}
	\p_x u_0=\frac{(u_1-u_0)(u_2-u_0)}{f},\ \ \p_x{u_1} = \frac{1}{u_2-u_0},\ \ \p_x u_2=\frac{(u_2-u_1)(u_2-u_0)}{f},
	\label{s1rr}
\end{equation}
where
\begin{equation}
	f=\big((u_2-u_1)(u_0-u_1)+\kappa\big)(u_{2}-u_0),
	\label{f}
\end{equation}
which can be interpreted as a differential constraint compatible with the potential dNLS system.
In order to verify this, one has to eliminate the $x-$derivatives in the two copies ($(u_0,u_1)$ and $(u_1,u_2)$) 
of the potential dNLS systems using (\ref{s1rr}), and check that derivatives $\p_t$ and $\p_x$ commute.
By construction, (\ref{s1rr}) is invariant with respect to the mapping defined
by equation (\ref{h1r3}):
\begin{equation}
M:
(u_{0}, u_{1},u_{2})\to
	\left(u_{1}, u_{2}, \ds u_0+\frac{\kappa}{u_2-u_1}\right).
	\label{h1rmap}
\end{equation}
This implies, in particular, that derivative $\p_x$ preserves the integral(s) of mapping $M$. 
One can check that $M$ has only one integral given by (\ref{f}) - it is also the integral of (\ref{s1rr}).
This integral can be obtained by means of the staircase method \cite{stair1,stair2} (see also \cite{stair3}).

\subsection*{Concluding remarks} 
Integrable quad-equations provide us with auto-transformations for solutions of some NLS-type 
systems. 
Although we used only one example of $H_1$ -- dNLS equations, the presented construction
is not specific to this case. It can be applied to other integrable quad-equations 
as well -- this will be the subject of further research.

\vspace{2mm}

The author is grateful to V.E. Adler and W.K. Schief for clarifying comments and indicating
some relevant references.
\vspace{5mm}

\end{document}